\documentclass{appolb}
\usepackage{graphicx}

\usepackage{amssymb}
\usepackage{amsmath}

\begin{document}
\title{A STATISTICAL METHOD FOR IMPROVING MOMENTUM MEASUREMENT OF PHOTON CONVERSIONS RECONSTRUCTED FROM SINGLE ELECTRONS}

\author{Ahmet Bing\"ul, Zekeriya Uysal
\address{Department of Engineering of Physics, Gaziantep University, Gaziantep 27310, Turkey.}
\\
}

\maketitle
\begin{abstract}
The reconstruction of photon conversions is importantin order to improve
the reconstruction efficiency of the physics measurements involving photons.
However, there are significant number of conversions in which only one of
the two tracks emitted electrons is reconstructed in the detector due to very asymmetric energy sharing
between the electron-positron pair.
The momentum determination of the parent photon can be improved
by estimating the missing energy in such conversions.
In this study, we propose a simple statistical method that can be used to determine the mean value of the missing energy.
By using simulated minimum bias events at LHC conditions and
a toy detector simulation, the performance of the method is tested
for several decay channels commonly used in particle physics analyses.
A considerable improvement in the mass reconstruction precision is obtained
when reconstructing particles decaying to photons whose energies are less than 20 GeV.
\end{abstract}
  
\section{Introduction}
In a photon conversion, usually an electron-positron pair is produced when a
photon of enough energy interacts with a nucleus.
In high-energy collision experiments, one has high photon multiplicity
in an event leading to production of many photon conversion candidates within the
detector material.
Therefore, the reconstruction of photon conversions in particle detectors
becomes important
for a variety of physics analysis involving electromagnetic decay products such as
$\pi^0 \rightarrow \gamma \gamma$.
The conversion reconstruction may also be used for detector-related
studies. For example, mapping the distribution of the conversion vertices allows
one to produce a
precise localization of the material in the tracker
detectors~\cite{CITE_ATLAS_CSC, CITE_ATLAS_PHT}.

Some fraction of the photon conversions will be highly asymmetric due to the
energy sharing mechanism between the products. Either the electron ($e^-$) or the positron
($e^+$) can be produced with a very low energy. If this energy falls below a threshold
required to produce a reconstructible track in the detector
then the converted photon will be seen to have only one track
and this type of conversion can be called a {\it single electron conversion}
or a {\it single track conversion}.
Even though there is a missing energy in such conversions,
one can still use them to increase the efficiency of the reconstructed photons.

The missing energy of the unreconstructed track cannot be predicted
kinematically since the parent photon energy is also unknown.
However, for a large collection of the reconstructed tracks,
the {\it mean} value of the missing energy can be evaluated
as a function of the reconstructed energy and pseudo-rapidity.
This paper describes a statistical procedure to estimate the missing energy
in a single track photon conversion.
By using the simulated minimum bias events at LHC conditions
and a suitable toy detector simulation, the performance of the method is examined
for the decay channels $\pi^0 \rightarrow \gamma \gamma$,
$\eta^0 \rightarrow \gamma \gamma$ and $D^{*0} \rightarrow D^0 \gamma$
and remarkable improvements in the mass reconstruction accuracies are observed.

The physics and reconstruction of 
photon conversion are given in
Section~\ref{SECTION_THEORY} and~\ref{SECTION_RECO} respectively.
Section~\ref{SECTION_SIMULATION} describes the event and detector simulation
selected for the study.
The statistical method is introduced in Section~\ref{SECTION_MEC}.
The performance of the method is presented in Section~\ref{SECTION_PERFORMANCE}.
Finally, a conclusion is given in Section~\ref{SECTION_CONCLUSION}.

\section{Physics of photon conversion}
\label{SECTION_THEORY}
At photon energies above 1 GeV, the interaction of the photons with a material
will be dominated by $e^- e^+$ pair production.
The photo-electric effect as well as the Rayleigh and Compton scattering cross sections are orders
of magnitude below that of the photon conversion.

A detailed description for the cross section of the photon conversion process
can be found in~\cite{CITE_TSAI} and in the GEANT4 Physics Reference
Manual~\cite{CITE_GEANT4}.
Here, a relatively simplified model is used as described in~\cite{CITE_KLEIN}.
Accordingly, for the photon energies used in this study
($E_\gamma>1$ GeV), the differential cross section can be approximated by:

\begin{equation}
\label{EQUATION_DIFF_CS}
  \frac{d\sigma}{dx} = \frac{A}{X_0 N_A} (1-\frac{4}{3}x(1-x))
\end{equation}
where $x = E_{e}/E_\gamma$ is known as the fractional electron energy or
the energy sharing, $A$ is the atomic mass in g/mol and $N_A$ is the
Avogadro's number.
$X_0$ is the radiation length in g/cm$^2$ along the path of the photon.
For a material whose atomic number is $Z$, $X_0$ can be evaluated from:

\begin{equation}
\label{EQUATION_X0}
  X_0 = \frac{716.4~A}{Z(Z+1)\ln(287 / \sqrt{Z})}
\end{equation}
One can show that the total cross section is independent of the incident photon
energy and given by:

\begin{equation}
\label{EQUATION_TOTAL_CS}
  \sigma_\text{tot} = \int_0^1{\frac{d\sigma}{dx}dx} = \frac{7A}{9 X_0 N_A}
\end{equation}
Also, a normalized distribution of $x$ can be obtained from:

\begin{equation}
\label{EQUATION_RHO}
  \rho(x) = \frac{1}{\sigma_{tot}}\frac{d\sigma}{dx} = \frac{9}{7}
(1-\frac{4}{3}x(1-x))
\end{equation}
Here $\rho(x)$ is symmetric function in $x$ and $1-x$, the electron and
positron energies respectively.
Since $x$ can have any random real value in the range $(0, 1)$,
the incident photon energy is not normally shared equally between the electron and
positron. Hence, the single track photon conversion may turn out
as a result of photons with a very asymmetric
energy sharing between the electron and the positron.

At the reconstruction level, a transverse momentum cut, $p_{T,\text{cut}}$, is
applied to form a list of {\it good} charged track candidates in an event.
For example, $p_T > 0.5$ GeV/c is required in many analyses at LHC conditions.
However, the use of $p_{T,\text{cut}}$ results in an angular dependence
in the reconstruction of the photon conversion process as follows.
The probability of observing a single track photon conversion can be defined as:

\begin{equation}
\label{EQUATION_PROB}
  P(x < x_t) =  \int_0^{x_t} \rho(x) dx
\end{equation}
where $x_t = p_{T,\text{cut}} \cosh \eta / E_\gamma$ is the threshold value of
the energy sharing and $\eta$ is the pseudo-rapidity of the parent
photon\footnote{In this calculation the mass of the electron is ignored.}.
It can be shown that the single track conversion occurs more likely
at lower photon energies and/or at higher $\eta$ values.

\section{Reconstruction of photon conversion}
\label{SECTION_RECO}
The signatures of charged particles are usually obtained
from the layered hit information in tracking detectors.
As an example, the ATLAS Inner Detector is a composite tracking system
consisting of pixels, silicon strips and straw tubes in a 2 T
magnetic field. The tracking procedure selects track candidates with
transverse momenta above 0.5 GeV/c.
As for electrons, their energies can be determined from the energy
deposition in the electromagnetic calorimeter. This measurement is combined
with the measurement of the electron momentum in the
inner detector to improve the energy
resolution~\cite{CITE_ATLAS_FIRSTDATA, CITE_ATLAS_TRACKING}.

In order to reconstruct the photon conversions a special vertex finding and
fitting algorithm is used. This algorithm combines oppositely charged tracks
in the event to evaluate the momenta of the conversion products at a vertex
point in the detector. Since the converted photon is a massless particle,
it also applies an additional angular constraint
such that the two tracks produced at the vertex should have an initial
difference of zero in their azimuthal and polar angles~\cite{CITE_CERN_OPEN}.
At the end of this procedure, one has a list of double track photon
conversion candidates in the event.

The remaining tracks can be re-examined to collect possible single track
conversions. In ATLAS, an electron-like track having its first hit beyond the pixel
vertexing layer is a well known signature of the surviving leg of the single track conversion.
Also, in rare cases, the conversion may happen so far from the beam axis that
the conversion pair can merge. If they do not traverse a long enough distance
inside the tracker, they cannot be resolved and as a result
a single track is reconstructed~\cite{CITE_ATLAS_CSC, CITE_CERN_OPEN}.
The other issue is to determine the production point of the single track
conversion. Actually, there is no way to know the precise position because it may occur
somewhere between detector layers (on the cables, on the cooling pipes etc).
By definition, one can set the position of the single conversion to the first
hit on the track.

%
\section{Event and detector simulation}
\label{SECTION_SIMULATION}
For the aim of the study the Pythia 8.1 tune 4cx
event generator~\cite{CITE_PYTHIA, CITE_TUNEX} under LHC conditions
(p-p collisions at the center of mass energy of 14 TeV) is
used to generate about 200,000 events with the minimum bias processes.
The simulated samples have additional collision events, pile-up,
added such that the average number of interactions per event is 22.
Also, the position of the interaction point is smeared around the origin
by Gaussian distributions in $x$, $y$ and $z$, such that the widths are
selected as $\sigma_x=\sigma_y = 0.05$ mm and $\sigma_z=60$ mm.

The selected events are passed through a toy detector simulation.
The momentum resolution is simulated by a single Gaussian smearing the
momentum of the charged tracks as follows:
\begin{equation}
\label{EQUATION_TOY_SIGMA_PT}
  \sigma_{p_T} / p_T = A p_T \oplus B
\end{equation}
where the transverse momentum, $p_T$, is in GeV/c and
$A$ and $B$ are numerical constants. In this study,
ATLAS detector parameterization where $A = 0.05\%$ and B = 1\% is
used~\cite{CITE_ATLAS_CSC}.
The reconstructed charged particles are required to have the
pseudo-rapidity $|\eta|<2.5$ and transverse momentum $p_T>0.5$ GeV/c.

\begin{figure*}[htb]
\begin{center}
\resizebox{0.6\textwidth}{!}{\includegraphics{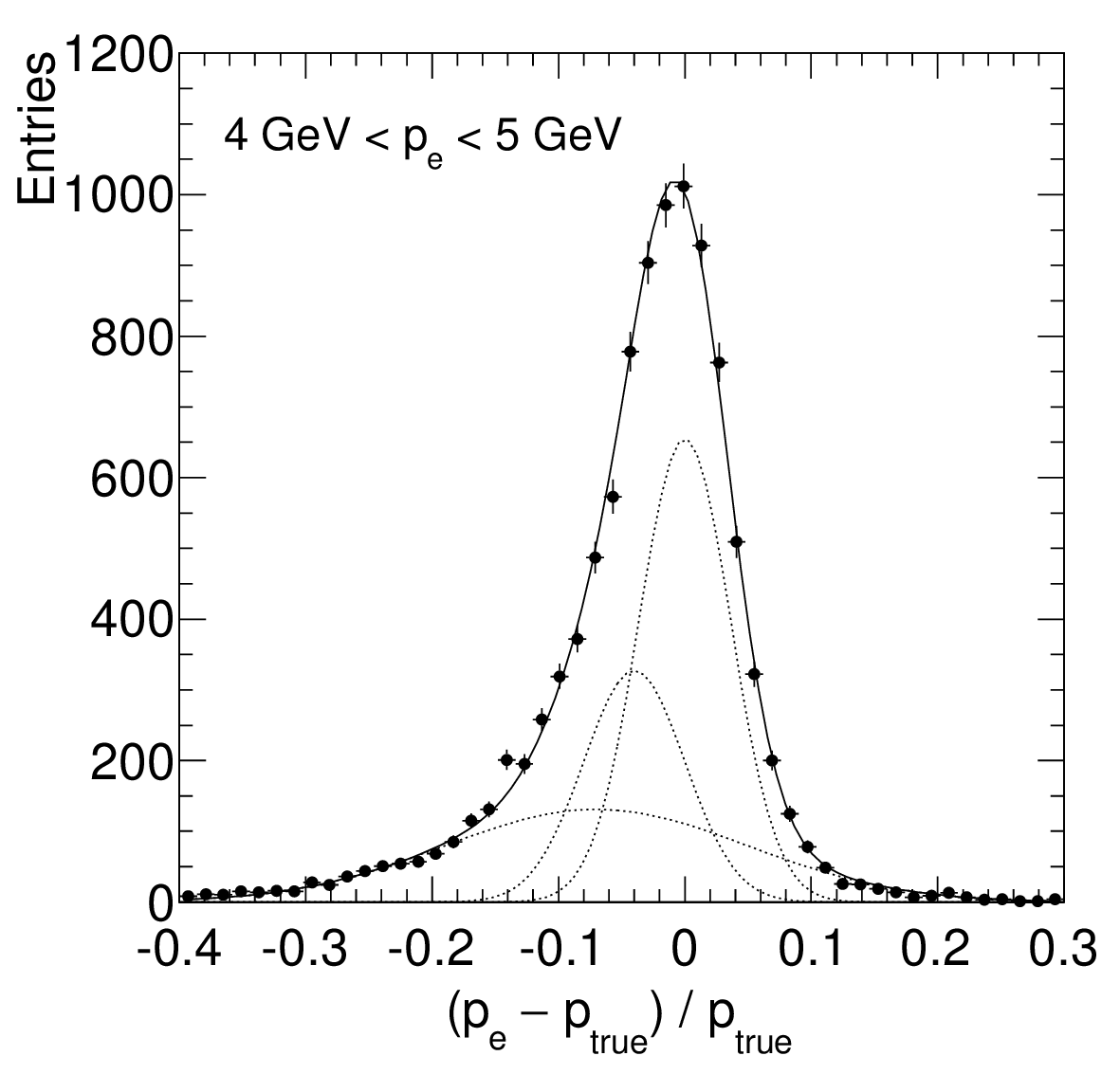}}
\caption{\label{FIGURE_GSF}
Example of the relative residual distribution of electrons
for a given momentum range and for all eta bins.
Data points are simulated with a sum of three normal distributions
with the first Gaussian function forced to describe the peak region
and the second and third Gaussian functions to model the tail to negative values.
}
\end{center}
\end{figure*}

The behavior of electrons in the detector is dominated by
the radiative energy losses (bremsstrahlung) as they traverse the matter.
Bremsstrahlung is a highly non-Gaussian process and cannot be adequately
represented by a single Gaussian function as in Equation~\ref{EQUATION_TOY_SIGMA_PT}.
In this study, the bremsstrahlung losses are modeled by a sum of
three Gaussians to smear the electrons' momenta in an event.
This idea is extracted from the Gaussian Sum Filter approach~\cite{CITE_GSF}
which is successfully applied to improve the track parameters of electrons both
in ATLAS~\cite{CITE_ATLAS_GSF} and in CMS~\cite{CITE_CMS_GSF} experiments.
Figure~\ref{FIGURE_GSF} shows an example of the relative residual distribution
of the electrons where $p_e$ is the reconstructed momentum and $p_{true}$
is the true momentum at the generator level.
The distribution has a Gaussian core and a large tail to negative values
caused by radiative energy losses of the electrons.
The mean values of the three Gaussian functions are required to be decreasing
such that the first Gaussian function always describes the peak region of the
distribution at around zero.

In the simulation, we have assumed that the tracking detector has
20 very long barrel layers which are regularly separated by 20 mm.
All photons with $E_\gamma > 1$ GeV in an event are converted at
a random position between $r = [20~\text{mm}, 440~\text{mm}]$
where $r$ is the conversion radius measured from the beam
axis\footnote{At LHC, the semi-diameter of the beam pipe is $r=20$ mm.}.
The cartesian coordinates of the conversion point can be found by extrapolating a
line from the true photon production point to a cylindrical surface of radius $r^\prime$
which is the radius of the closest detector layer
just after the conversion point along the path of the photon.
Finally, the direction of the conversion products
are assumed to be the same as the parent photon.

%

\section{The statistical correction method}
\label{SECTION_MEC}
In this section, the statistical method for compensating the missing energy
in a single track conversion is discussed. The mass of the electron and
the recoiling kinetic energy of the nucleus are neglected in the calculations.

In the single track conversion, the conservation of energy yields:

\begin{equation}
\label{EQUATION_MEC1}
E_\gamma = E_\text{reco} + E_\text{miss}
\end{equation}
where $E_\text{reco}$ is the reconstructed energy after applying
the momentum smearing and transverse momentum cut and
$E_\text{miss}$ is missing energy of the unreconstructed track at the truth level.
In the most simple case one can assume that $E_\text{miss}=0$, i.e. the photon
energy is carried only by the surviving track.

In fact, the missing energy can be estimated by a Monte Carlo study.
For a given $p_{T\text{cut}}$, $E_\text{miss}$ will be distributed in
the range $(0,~p_{T\text{cut}} \cosh \eta)$.
Using a large enough number of minimum bias events\footnote{At LHC conditions,
a few million minimum bias events can be collected in a few hours.}, one can
extract the mean value of the missing energy distribution,
$\langle E_\text{miss} \rangle$, which is expected to be a function of the
energy and pseudo-rapidity of the reconstructed track.
Replacing $E_\text{miss}$ by $\langle E_\text{miss} \rangle$
in Equation~\ref{EQUATION_MEC1} improves on average the value of $E_\gamma$.

\begin{figure*}[!htb]
\begin{center}
\resizebox{0.6\textwidth}{!}{\includegraphics{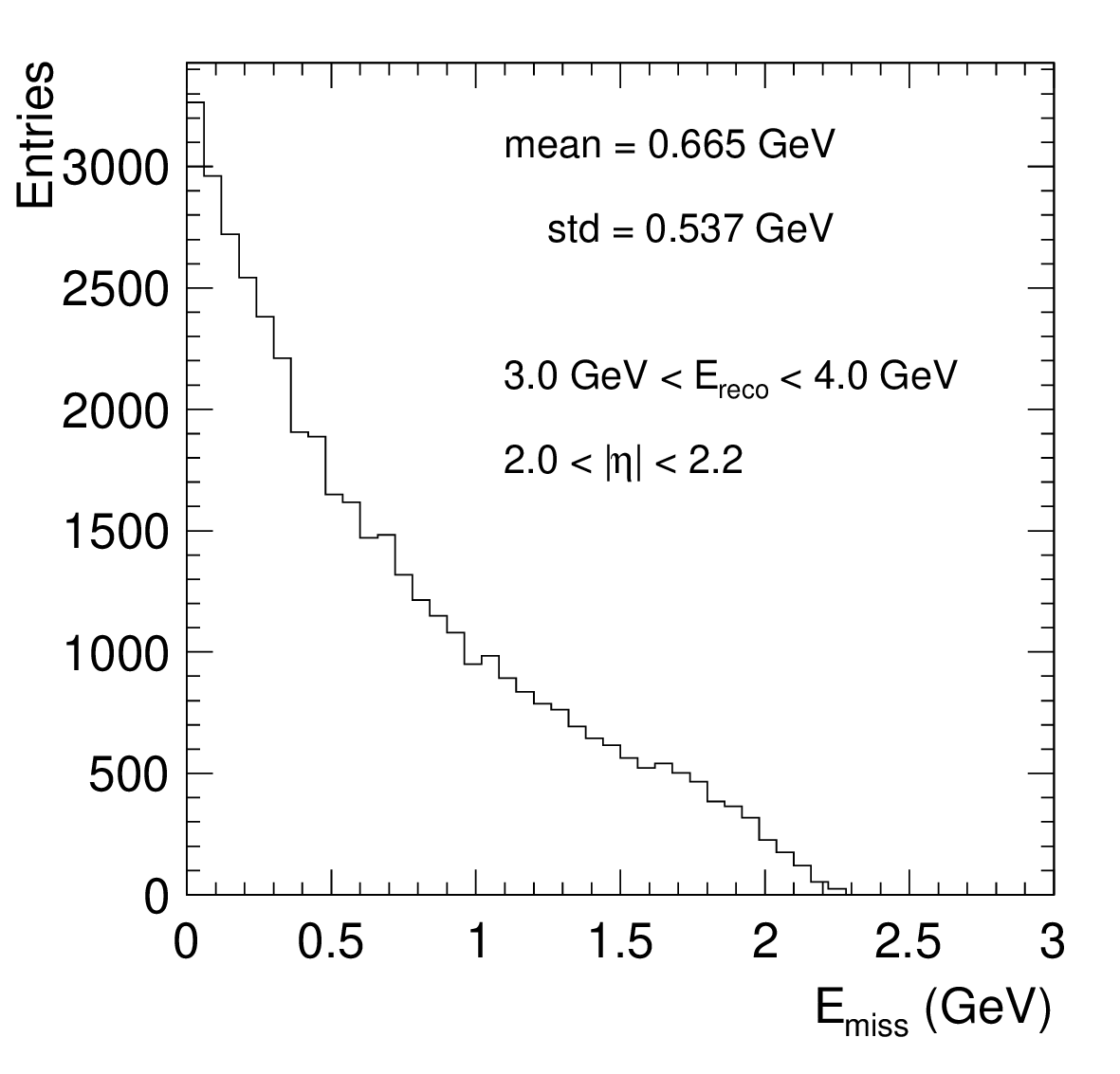}}
\caption{\label{FIGURE_MISSEN}
The missing energy distribution obtained from the Monte Carlo truth information
for the specific reconstructed energy and pseudo-rapidity ranges.
}
\end{center}
\end{figure*}

Figure~\ref{FIGURE_MISSEN} shows an example missing energy distribution
for $3.0~\text{GeV}<E_\text{reco}<4.0~\text{GeV}$, $2.0<|\eta|<2.2$ and
$p_{T,\text{cut}} = 0.5$ GeV/c$^2$.
For these intervals, $\langle E_\text{miss} \rangle = 0.665$ GeV and therefore
the photon energy can be calculated simply from
$E_\gamma = E_\text{reco} + 0.665$ GeV.

\begin{figure*}[htb]
\begin{center}
\resizebox{0.7\textwidth}{!}{\includegraphics{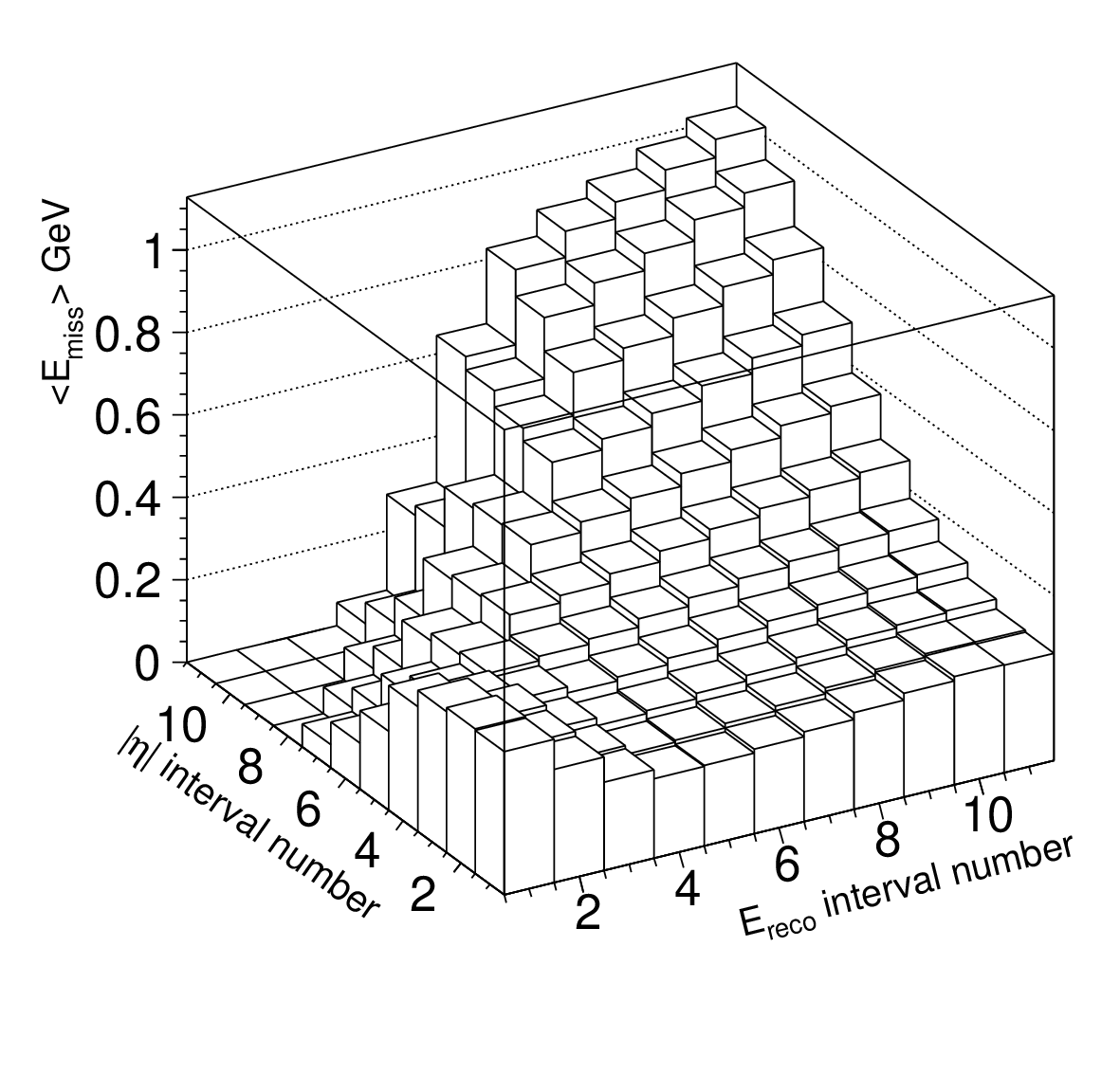}}
\vspace{-5mm}
\caption{\label{FIGURE_MATRIX}
The lego plot indicating the values of the missing energy matrix elements ($\langle E_\text{miss} \rangle$)
obtained from the Monte Carlo truth study as a function of the reconstructed energy
and pseudo-rapidity interval numbers given in Table~\ref{TABLE:INTERVALS}.
There are relatively larger fluctuations in lower energy and pseudo-rapidity intervals.
}
\end{center}
\end{figure*}

In the study, 11 energy and 11 pseudo-rapidity intervals, given in
Table~\ref{TABLE:INTERVALS}, are
defined\footnote{Selection of the intervals depends on the statistics and data used.}
for all single track conversions whose parent photons originate from any
mother particle. A lego plot based on the $E_\text{reco}$ vs $\eta$ interval numbers,
is shown in Figure~\ref{FIGURE_MATRIX} where the height of each bin stands for
the mean value of the missing energy.
These mean values are stored in a matrix for use in a physics analysis later.
It is found that the values of the matrix elements depend on the detector resolutions
and the kinematic cuts applied for selecting particles.

\begin{table}[!htp]
\begin{center}
 \label{TABLE:INTERVALS}
 \caption{The energy and pseudo-rapidity intervals defined for
          the reconstructed tracks in the conversions.}
  \begin{tabular}{ c l l}
   \hline
    Interval     & Energy (GeV)               & Pseudo-rapidity    \\
    \hline
    1            & $0.50<E_\text{reco}<0.75$  & $|\eta|<0.25$      \\
    2            & $0.75<E_\text{reco}<1.00$  & $0.25<|\eta|<0.50$ \\
    3            & $1.00<E_\text{reco}<1.25$  & $0.50<|\eta|<0.75$ \\
    4            & $1.25<E_\text{reco}<1.50$  & $0.75<|\eta|<1.00$ \\
    5            & $1.50<E_\text{reco}<2.00$  & $1.00<|\eta|<1.25$ \\
    6            & $2.00<E_\text{reco}<3.00$  & $1.25<|\eta|<1.50$ \\
    7            & $3.00<E_\text{reco}<4.00$  & $1.50<|\eta|<1.75$ \\
    8            & $4.00<E_\text{reco}<5.00$  & $1.75<|\eta|<2.00$ \\
    9            & $5.00<E_\text{reco}<7.50$  & $2.00<|\eta|<2.20$ \\
    10           & $7.50<E_\text{reco}<10.0$  & $2.20<|\eta|<2.40$ \\
    11           & $10.0<E_\text{reco}<20.0$  & $2.40<|\eta|<2.50$ \\
   \hline
 \end{tabular}
\end{center}
\end{table}

The final issue is to determine the direction of the parent photon which is
defined in two ways. First, it can be considered as having the same direction of
the reconstructed track. In this case, the four-vector of the photon can
be defined as:

\begin{equation}
\label{EQUATION_MEC2}
P_\gamma = (E_\gamma \frac{p_{x,\text{reco}}} {p_{\text{reco}}},~
            E_\gamma \frac{p_{y,\text{reco}}} {p_{\text{reco}}},~
            E_\gamma \frac{p_{z,\text{reco}}} {p_{\text{reco}}},~
            E_\gamma)
\end{equation}
where $p_{x,y,z,\text{reco}}$ are the xyz-components of the momentum at the
production point and $p_\text{reco} \approx E_\text{reco}$ is the magnitude
of the momentum of the reconstructed track.
Second, one can assume that the photon is originating from the primary vertex
in the event. Then, the photon will be in the direction of the line joining the primary
vertex $A = (x_0,~y_0,~z_0)$ to the pair production point $B = (x_p,~y_p,~z_p)$.
In our toy simulation, we set $A = (0,~0,~0)$. This approach yields a four-vector:

\begin{equation}
\label{EQUATION_MEC3}
P_\gamma = (E_\gamma \frac{x_p}{R_p},~E_\gamma \frac{y_p}{R_p},~E_\gamma
\frac{z_p}{R_p},~E_\gamma)
\end{equation}
where $R_p = |AB| = (x_p^2+y_p^2+z_p^2)^{1/2}$.
No significant difference is observed between the use of
the equations ~\ref{EQUATION_MEC2} and~\ref{EQUATION_MEC3} in the
simulation.

To avoid biases in the training data (over-training),
the missing energy matrix is produced by using a training data set
which is the first half of the full data.
The validation of the method is performed using an independent test data
which is the rest of the full data. The use of single track
photon conversions in our analysis is summarized in the following algorithm.

\begin{enumerate}
 \item Generate an event.
 \item Select a photon from any source in the event.
 \item Convert the photon at a random point in the detector.
 \item Obtain the energy of the electron and positron from the distribution
       given in Equation~\ref{EQUATION_RHO}.
 \item Smear the momentum of the electron and positron and determine
       their four-vectors whose space coordinates are in the direction of the photon.
 \item If one of the energy of the conversion products falls below the threshold,
       evaluate the value of the reconstructed energy, $E_\text{reco}$, and
       the value of $\langle E_\text{miss} \rangle$ from the mean missing energy matrix
       depending on $E_\text{reco}$ and $\eta$.
 \item Compute the energy of the reconstructed photon from
       $E_\gamma = E_\text{reco} + \langle E_\text{miss} \rangle$ and
       evaluate its four-vector from Equation~\ref{EQUATION_MEC2}.
 \item Use the reconstructed photon in the analysis.
\end{enumerate}

\section{Performance}
\label{SECTION_PERFORMANCE}
The performance of the statistical procedure given in the
previous section is tested for three different decay channels.
The nominal mass of each particle is taken from Ref~\cite{CITE_PDG}
and, the transverse momentum cut is taken as $p_{T,\text{cut}} = 0.5$ GeV/c
unless otherwise stated.

\subsection{Invariant mass distributions}
\begin{figure*}[htb]
\begin{center}
\resizebox{0.7\textwidth}{!}{\includegraphics{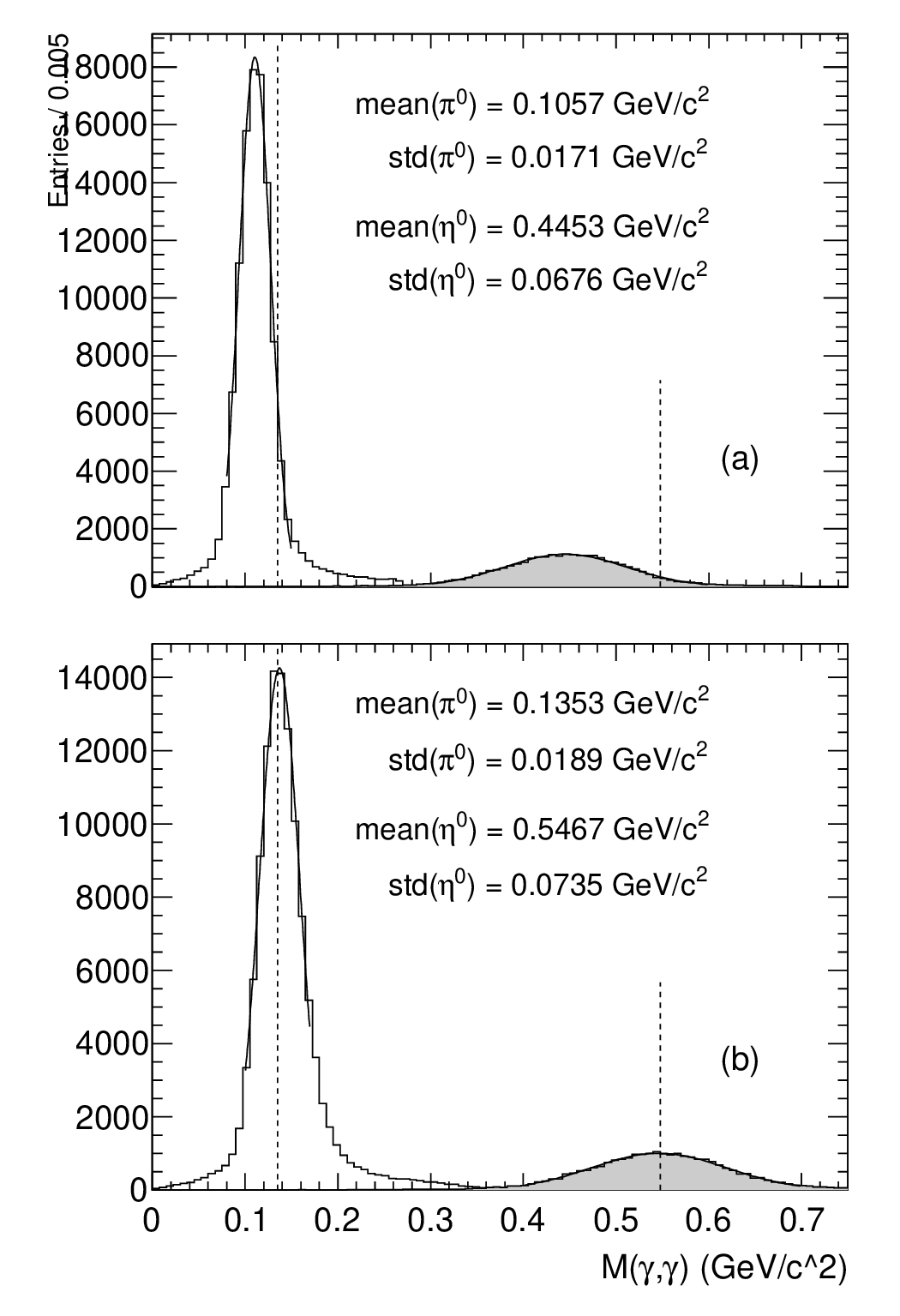}}
\caption{\label{FIGURE_MGG}
Two-photon invariant mass spectra for $\pi^0$ and $\eta^0$ signals,
where both photons are reconstructed from the single track conversion,
when (a) $E_\text{miss}=0$ and (b) $E_\text{miss}=\langle E_\text{miss} \rangle$.
The nominal mass positions, $M_{\pi^0} = 0.1349766$ GeV/c$^2$ and
$M_{\eta^0} = 0.547862$ GeV/c$^2$, are indicated by the dashed lines.
mean and std represent respectively the center and the width of the Gaussian fits.
}
\end{center}
\end{figure*}

Firstly, the common decays containing two photons,
$\pi^0 \rightarrow \gamma \gamma$ and $\eta^0 \rightarrow \gamma \gamma$,
are considered. For the reconstructed photon energies $E_1$ and $E_2$,
the two-photon invariant mass is defined by:

\begin{equation}
\label{EQUATION_MGG}
M_{\gamma\gamma} = \sqrt{ 2 E_1 E_2 (1 - \cos \theta) }
\end{equation}
where $\theta$ is the opening angle between photons.
Figure~\ref{FIGURE_MGG}a and~\ref{FIGURE_MGG}b respectively show
$M_{\gamma\gamma}$ distributions of $\pi^0$ and $\eta^0$ signals
for $E_\text{miss} = 0$ and $E_\text{miss} = \langle E_\text{miss} \rangle$
where both photons are built from the single track conversion.
The photon energies, $E_1$ and $E_2$, tend to have lower values
when $E_\text{miss} = 0$. As a result, $\pi^0$ and $\eta^0$ signals are
shifted to lower mass values according to Equation~\ref{EQUATION_MGG}.
However, by correcting the missing energy, the mean of the distributions
are moved to their nominal positions as indicated by the dashed lines on the figures.
The peak regions of both signals are fitted to Gaussian functions.
The center positions and widths of the fits are shown as mean and std respectively
on the figure.

\begin{figure*}[htb]
\begin{center}
\resizebox{0.7\textwidth}{!}{\includegraphics{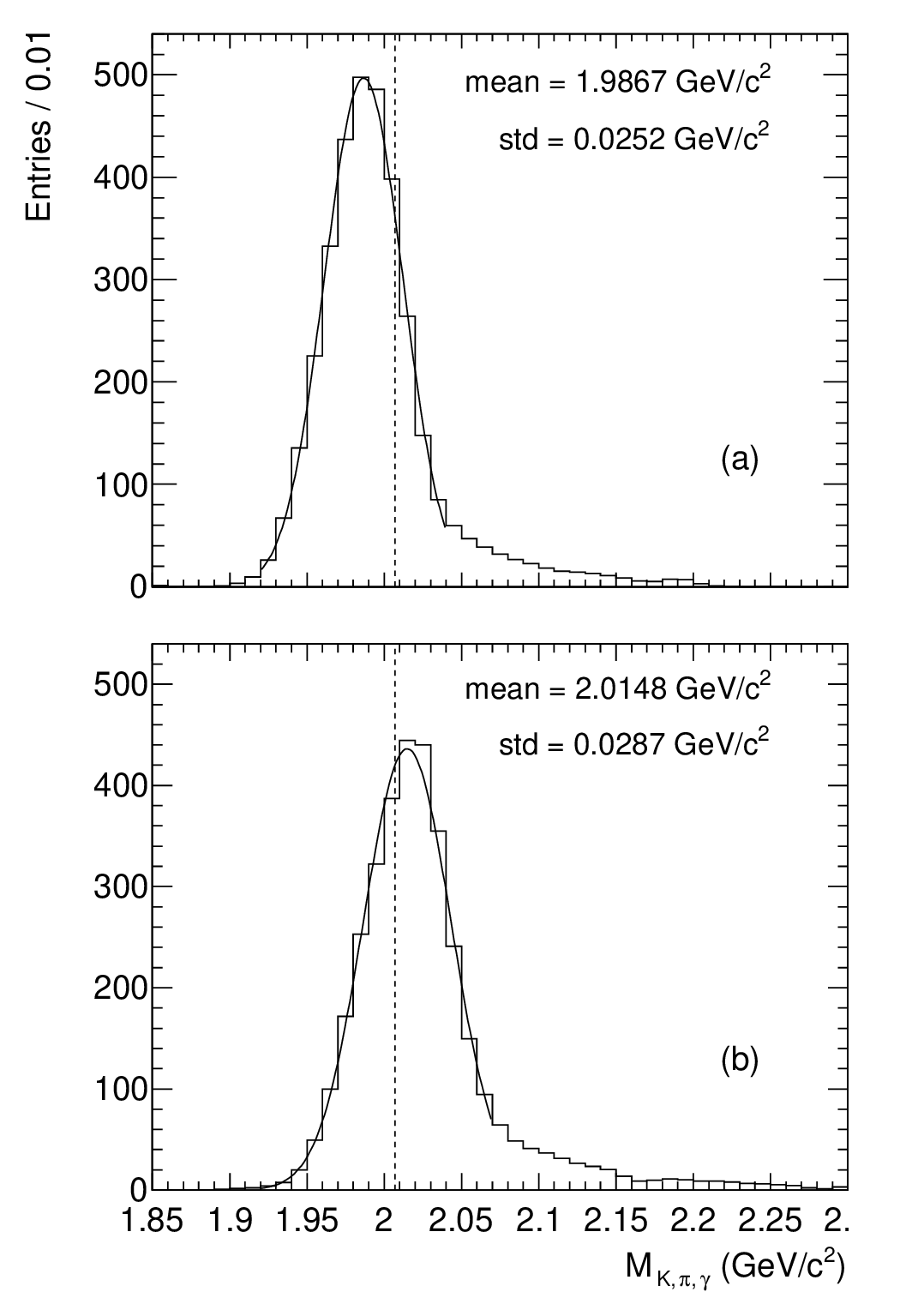}}
\caption{\label{FIGURE_MDS}
$M_{K~\pi,~\gamma}$ spectra for $D^{*0}$ signal
where the photon is reconstructed from the single track conversion,
when (a) $E_\text{miss}=0$ and (b) $E_\text{miss}=\langle E_\text{miss} \rangle$.
The nominal mass position, $M_{D^{*0}} = 2.00699$ GeV/c$^2$, is indicated by a dashed line.
mean and std represent respectively the center and the width of the Gaussian fits.
}
\end{center}
\end{figure*}

The same procedure is repeated for the decay $D^{*0} \rightarrow D^0 \gamma$
where $D^0$ candidates are selected from the channel $D^0 \rightarrow K^\pm
\pi^\mp$. The three-particle invariant mass, $M_{K,~\pi,~\gamma}$, distributions
before and after correction are shown in Figure~\ref{FIGURE_MDS}a and
~\ref{FIGURE_MDS}b respectively.
It is clear that the mean correction method improves the accuracy of the
$D^{*0}$ mass reconstruction.


\subsection{Energy resolution}
\begin{figure*}[htb]
\begin{center}
\resizebox{0.6\textwidth}{!}{\includegraphics{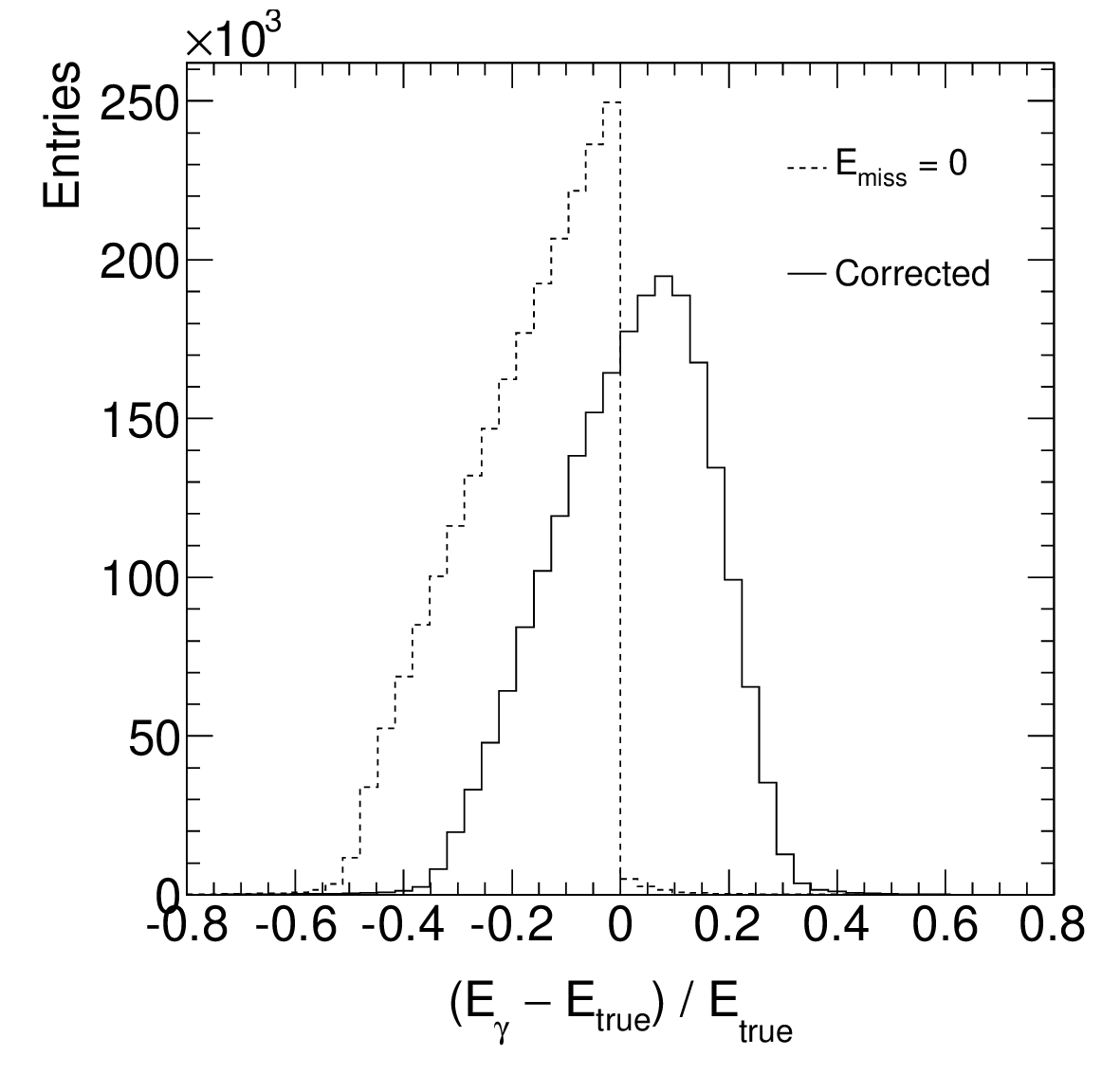}}
\caption{\label{FIGURE_RES}
The relative residual distributions of photons built from single track conversions
for all photon energy and pseudo-rapidity intervals.
}
\end{center}
\end{figure*}

The energy resolution of the photons can be extracted from the standard
deviation of the relative residual distribution defined by $(E_\gamma - E_{true}) / E_{true}$
where $E_{true}$ is the true photon energy at the generator level.
Figure~\ref{FIGURE_RES} shows the relative residual distribution
of photons reconstructed from single track conversions before and after
missing energy correction.
Clearly, the distribution is asymmetric with negative mean value when no
correction is applied. However, the proposed method shifts the mean of
the distribution to around zero while its width is slightly wider.

Similar relative residual distributions can be obtained for
the momenta of parent particles decaying to photons built from single track
conversions. However, further improvement in the parent's momentum can be obtained
by applying a mass constraint as described in~\cite{CITE_MASSCON} for example.

\subsection{Size of energy and $\eta$ bins}
In order to investigate how much the results depend on the binning choice,
coarser bins including only 6 energy and 6 pseudo-rapidity intervals are
defined such that the range of each interval in Table~\ref{TABLE:INTERVALS}
is doubled.

\begin{figure*}[htb]
\begin{center}
\resizebox{0.6\textwidth}{!}{\includegraphics{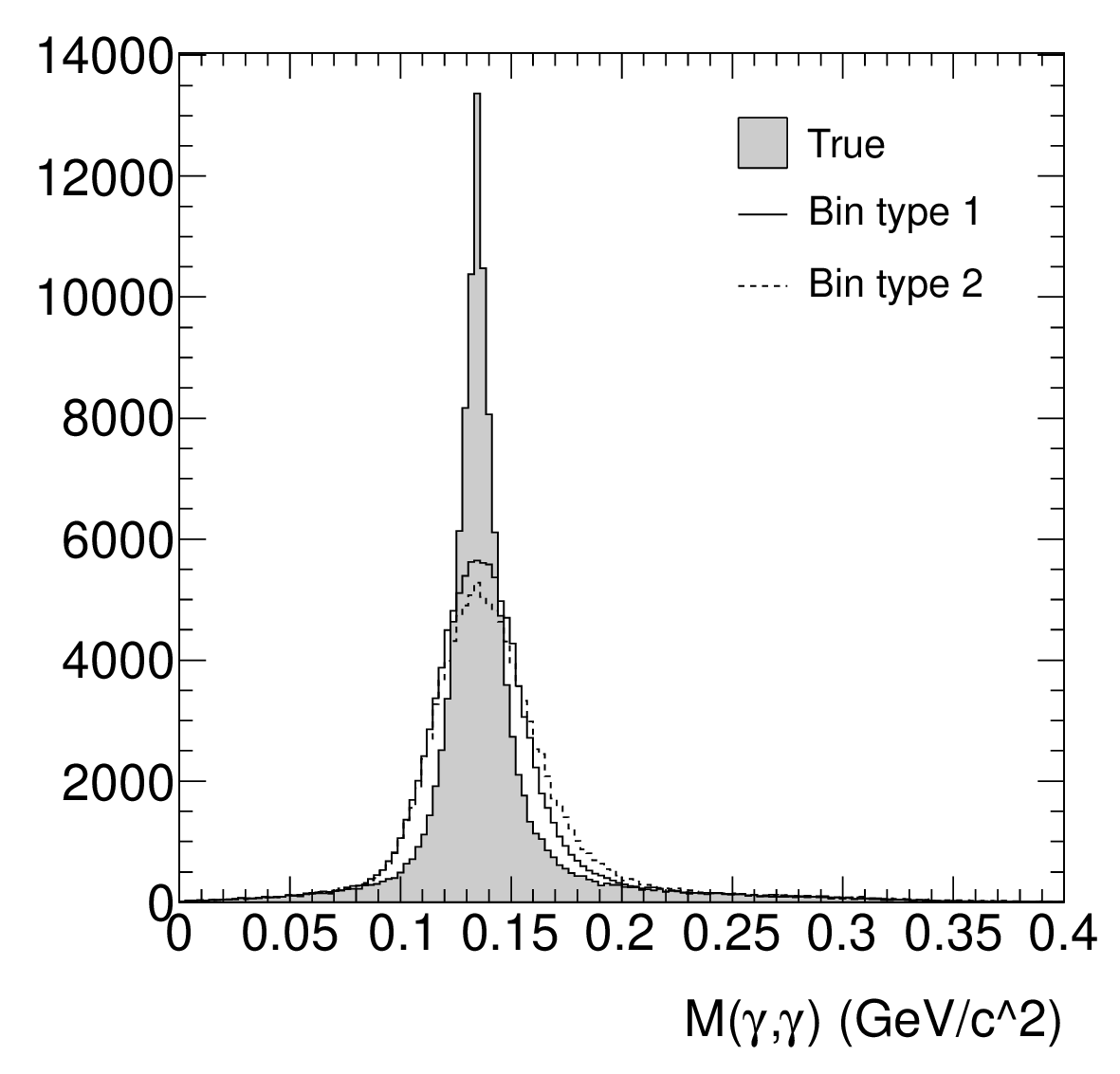}}
\caption{\label{FIGURE_BIN}
The effect of the selection of bin size on $\pi^0$ signal.
bin 1 and bin 2 are respectively represent intervals defined in Table~\ref{TABLE:INTERVALS}
and the other intervals defined in the text.
The shaded histogram is obtained when $E_\text{miss} = E_\text{true}$.
}
\end{center}
\end{figure*}

Figure~\ref{FIGURE_BIN} shows three $\pi^0 \rightarrow \gamma \gamma$
invariant mass distributions where both photons are reconstructed from
single track conversions. The solid and dashed histograms are respectively obtained
for the case $E_\text{miss} = \langle E_\text{miss} \rangle$ where
bin type 1 represents the binning in Table~\ref{TABLE:INTERVALS},
while bin type 2 represents the new energy and pseudo-rapidity bins.
The coarser bin selection result in a slightly wider mass distribution.
Hence, depending on the size of data, the selection of finer bins
turns out a better mass reconstruction precision.

In addition, to compare the results with the best reconstruction scenario,
one can choose $E_\text{miss} = E_\text{true}$. This is also shown as
a shaded histogram in the same figure. Note that, in the perfect case,
the true $\pi^0$ signal has still a significant mass resolution \textemdash FWHM of
the perfect distribution is about 4 times narrower than the bin type 1.
This is because the mass resolution is also affected by the
opening angle resolution between the photons arising from the four-vector
definition of single track conversions.

\subsection{Effect of $p_{T,\text{cut}}$}
\begin{figure*}[htb]
\begin{center}
\resizebox{0.7\textwidth}{!}{\includegraphics{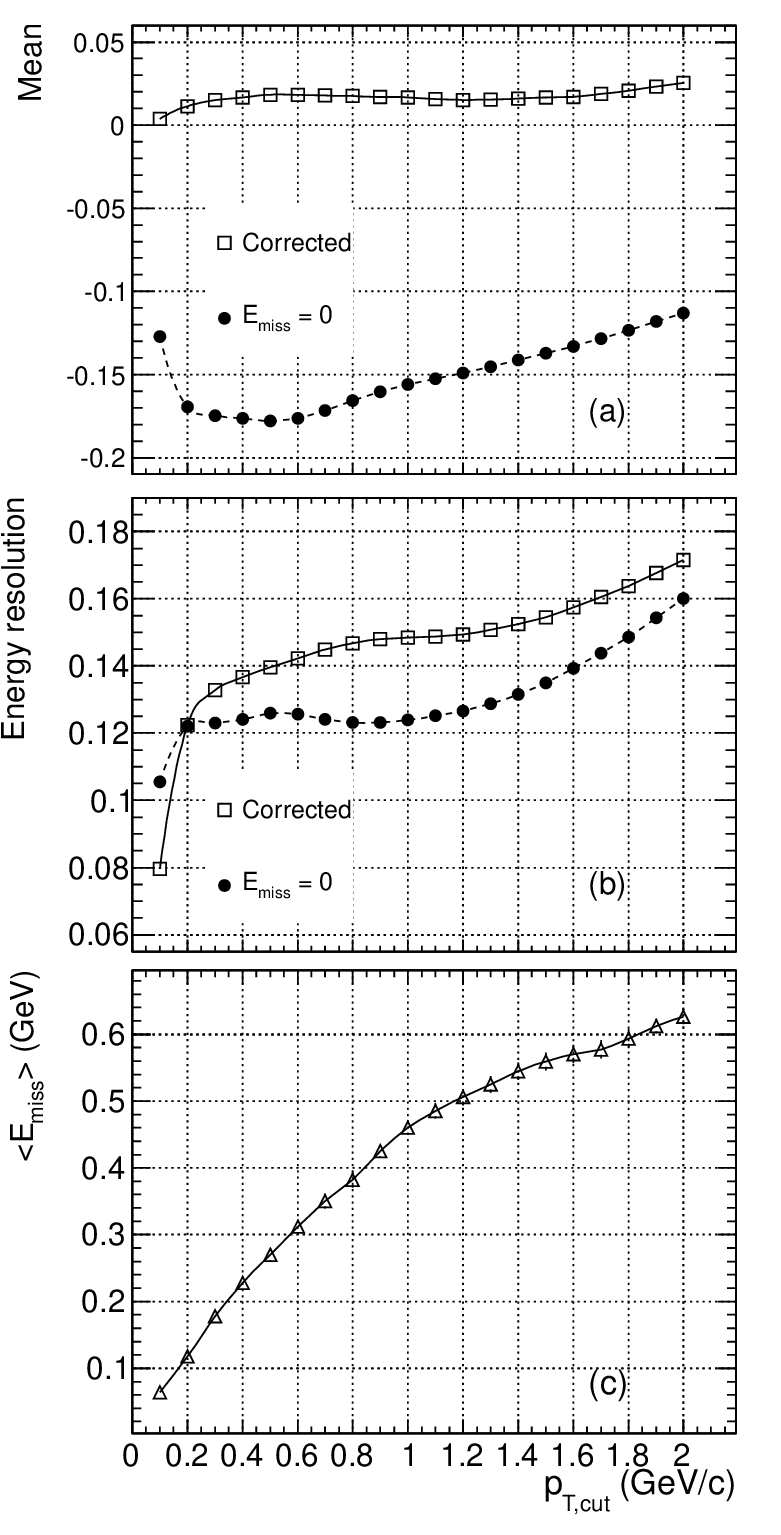}}
\caption{\label{FIGURE_RESPT}
(a) The mean value and (b) the energy resolution extracted from
the relative residual distributions as a function of $p_{T,\text{cut}}$.
(c) The mean values of the missing energy for a specific energy and pseudo-rapidity intervals.
for the same $p_{T,\text{cut}}$ values.
Error bars are much less than the marker sizes.
}
\end{center}
\end{figure*}

It is interesting to investigate the effect of $p_{T,\text{cut}}$
on the mean value and the energy resolution (width)  of the residual distributions.
As indicated in Figure~\ref{FIGURE_RESPT}a,
the mean values obtained after the correction is
slightly influenced by $p_{T,\text{cut}}$ and their values are close to zero.
As we expect, the mean values are always negative and far from zero
when no correction is applied.

Figure~\ref{FIGURE_RESPT}b shows the energy resolutions as a function of $p_{T,\text{cut}}$.
Clearly, the resolutions are getting worse with increasing $p_{T,\text{cut}}$.
A similar behaviour can be seen in Figure~\ref{FIGURE_RESPT}c which shows the
evolution of $\langle E_\text{miss} \rangle$ for a specific $E_\text{reco}$ and $\eta$
bins\footnote{The same trends are observed for all $E_\text{reco}$ and $\eta$ bins.}.
as a function of $p_{T,\text{cut}}$ up to 2 GeV/c.
Therefore, in consideration of $p_{T,\text{cut}}$, the strong positive correlation
between the energy resolution and $\langle E_\text{miss} \rangle$
points out that the overshooting in the energy correction of individual
photons results in the loss in the energy resolution of single track conversions.

%

\section{Conclusion}
\label{SECTION_CONCLUSION}
In a photon conversion one of the electron track can be missed for various reasons.
However, the reconstruction of such single-track photon conversions may be
a major concern for a number of physics analyses involving photons especially at the LHC.
In this study, a simple statistical method based on the average energy correction for
determining the missing energy in single-track conversions has been
introduced. By using the minimum bias events generated at LHC conditions
and a toy detector simulation, the method is shown to perform well when
it is applied to signal events like $\pi^0 \rightarrow \gamma \gamma$ and $D^{*0}
\rightarrow D^0 \gamma$ where the resulting photons are built from single track conversions.
It is found that the energy resolution of such photons are becoming worse
with increasing the value of $p_{T,\text{cut}}$.
However, the performance of the method may be improved by using finer energy and pseudo-rapidity bins.

The procedure described here can be implemented to improve the measurements and
discovery potentials of rare decays such as
$\Omega^- \rightarrow \Xi^- \gamma$ or $\Sigma^0 \rightarrow \Lambda \gamma
\gamma$.
However, in a real analysis, the reader must also consider the background
events acting as  single track or double track photon conversions for instance
$\pi^0 \rightarrow e^- e^+ \gamma$,
$K^0_S \rightarrow \pi^+ \pi^-$,
$\Lambda \rightarrow p \pi$,
$\Sigma^+ \rightarrow p \gamma$  etc.

The statistical approach is clearly not useful for correcting individual photon momenta.
However, the single-track photon conversions will still be highly desirable to improve
the reconstruction efficiency of particles decaying to photons in an event.
Therefore, a statistical approach similar to this study
might be very useful for scientists working with photon data
collected from experiments containing a high radiation length in the tracking
detectors.




\end{document}